\documentclass[conference]{IEEEtran}
\usepackage{amsmath}
\usepackage{graphicx}
\usepackage{cite}
 \usepackage{amsfonts}
\usepackage{algorithmic}
\usepackage{algorithm}
 \usepackage{multirow}
\usepackage{subcaption}

\usepackage[table]{xcolor}
\usepackage[graphicx]{realboxes}
\usepackage{titlesec}
\makeatletter
\makeatletter
\def\ps@IEEEtitlepagestyle{%
  \def\@oddfoot{\mycopyrightnotice}%
  \def\@evenfoot{}%
}
\def\mycopyrightnotice{%
  \gdef\mycopyrightnotice{}% just in case
}
\renewcommand\IEEEkeywordsname{Keywords}
\makeatother
\title{
APT-LLM: Embedding-Based Anomaly Detection of Cyber Advanced Persistent Threats Using Large Language Models
}

\DeclareRobustCommand{\IEEEauthorrefmark}[1]{\smash{\textsuperscript{\footnotesize #1}}}
\author{
    \IEEEauthorblockN{
        Sidahmed Benabderrahmane\IEEEauthorrefmark{1}, 
        Petko Valtchev\IEEEauthorrefmark{2}, 
        James Cheney\IEEEauthorrefmark{3}, 
        Talal Rahwan\IEEEauthorrefmark{1}
    } \vspace{-1mm}\\
    \IEEEauthorblockA{\IEEEauthorrefmark{1} New York University, Division of Science.} \\
    \IEEEauthorblockA{\IEEEauthorrefmark{2} University of Montreal, UQAM.}  \\
    \IEEEauthorblockA{\IEEEauthorrefmark{3} The University of Edinburgh, School of Informatics.}  \\
    \IEEEauthorblockA{\{sidahmed.benabderrahmane, talal.rahwan\}@nyu.edu, valtchev.petko@uqam.ca, jcheney@inf.ed.ac.uk} \\
}
\usepackage[justification=raggedright,singlelinecheck=false]{caption}

\begin{document}

\maketitle
\begin{abstract} 
Advanced Persistent Threats (APTs) pose a major cybersecurity challenge due to their stealth and ability to mimic normal system behavior, making detection particularly difficult in highly imbalanced datasets. Traditional anomaly detection methods struggle to effectively differentiate APT-related activities from benign processes, limiting their applicability in real-world scenarios.
This paper introduces APT-LLM, a novel embedding-based anomaly detection framework that integrates large language models (LLMs)—BERT, ALBERT, DistilBERT, and RoBERTa—with autoencoder architectures to detect APTs. Unlike prior approaches, which rely on manually engineered features or conventional anomaly detection models, APT-LLM leverages LLMs to encode process-action provenance traces into semantically rich embeddings, capturing nuanced behavioral patterns. These embeddings are analyzed using three autoencoder architectures—Baseline Autoencoder (AE), Variational Autoencoder (VAE), and Denoising Autoencoder (DAE)—to model normal process behavior and identify anomalies. The best-performing model is selected for comparison against traditional methods.
The framework is evaluated on real-world, highly imbalanced provenance trace datasets from the DARPA Transparent Computing program, where APT-like attacks constitute as little as 0.004\% of the data across multiple operating systems (Android, Linux, BSD, and Windows) and attack scenarios. Results demonstrate that APT-LLM significantly improves detection performance under extreme imbalance conditions, outperforming existing anomaly detection methods and highlighting the effectiveness of LLM-based feature extraction in cybersecurity.
\end{abstract}
\section{Introduction}
Advanced Persistent Threats (APTs) are a class of cyberattacks characterized by their stealth, sophistication, and long-term presence within targeted networks \cite{ahmad2019strategically}. Rather than executing quick, indiscriminate hits, APT campaigns carefully exploit system weaknesses, maintain persistence, and exfiltrate sensitive data over extended periods. Their persistent nature and ability to evade traditional detection methods—often by blending in with legitimate system processes—make APTs particularly challenging to identify and mitigate\cite{che2024systematic}.

In recent years, Pre-trained Large Language Models (LLMs) have emerged as a groundbreaking approach across numerous domains \cite{chang2024survey}. LLMs such as BERT, RoBERTa, GPT, or recently DeepSeek utilize enormous amounts of unlabeled text to learn rich contextual embeddings, enabling them to capture semantic nuances far beyond the capabilities of earlier, task-specific models. This success in natural language processing has inspired researchers to leverage LLM-based embeddings for novel applications—including cybersecurity—where the ability to represent complex “behavioral signatures” or “event sequences” as textual embeddings can potentially detect malicious patterns \cite{motlagh2024large}.

Building on these advances, we propose an LLM-driven APT detection framework designed to enhance the identification of stealthy behaviors within system logs and network traces. Our core idea is to convert low-level process events (e.g., file operations, network connections) into descriptive textual phrases, then extract embeddings using state-of-the-art LLMs. By combining these semantic-rich embeddings with an anomaly detection model—such as an autoencoder—we aim to detect hidden APT attacks in traces more effectively than conventional, signature-based methods. We evaluate our approach on real-world APT datasets, highlighting the potential of LLM embeddings to uncover subtle and malicious sequences often overlooked by standard methods.
\section{Related work}
APT detection has long relied on a combination of rule-based approaches, signature matching, and heuristic analysis \cite{che2024systematic}. Classical solutions frequently inspect network traffic, system calls, or event logs to identify suspicious patterns. However, attackers often disguise APT behavior within normal operating system activities, rendering purely signature-based methods insufficient. In the context of stealthy campaigns, conventional detection frameworks often fail to capture low-frequency or polymorphic events leading to data exfiltration. Recent works have introduced machine learning techniques—ranging from supervised classifiers to anomaly detection algorithms—aimed at capturing novel or stealthy behaviors \cite{koufakou_2007},\cite{smets2011},\cite{narita_2008},\cite{BerradaCBMMTW20},\cite{Benabderrahmane21},\cite{DBLP:journals/fgcs/BenabderrahmaneHVCR24}. For example, intrusion detection systems (IDS) increasingly adopt textual, time series or graph-based models that encode dependencies between system events. Despite these advancements, the sophistication of APT tactics continues to outpace purely traditional approaches.\\
In recent years, pre-trained LLMs have demonstrated remarkable capability in capturing semantic and contextual nuances across vast domains. Although they have primarily excelled in natural language processing tasks such as translation, summarization, and question-answering, researchers have begun to apply LLM-based embeddings to cybersecurity problems as well \cite{chen2024survey}. By transforming logs or system traces into textual narratives, LLMs can learn richer representations of process behaviors. These embeddings allow anomaly detection algorithms to isolate subtle deviations indicative of malicious activity.
Despite their potential, these methods are still in the early stages, facing challenges such as high computational cost and the need for extensive domain adaptation \cite{bayer2024cysecbert}.\\
Although LLMs have shown promise in detecting anomalies, most studies remain in the proof-of-concept stage, lacking large-scale evaluations on real data or heterogeneous operating systems (OS). In addition, limited attention has been paid to the interpretability of LLM-derived features, which are critical in high-stakes environments, where security analysts need clear rationales for flagged anomalies. To address these gaps, our work proposes a comprehensive pipeline for APT detection that fuses LLM-based embeddings with robust anomaly detection, systematically evaluated across multiple OS datasets and real-world threat scenarios.
\section{Proposed Framework}
\label{sec:framework}
In this section, we present our \emph{APT-LLM} detection framework, which leverages a provenance database of process events and netflow activities, pre-trained Large Language Models (LLMs) for embedding generation, and three autoencoder architectures for anomaly detection. The following paragraphs include key details to clarify each stage of our approach.
\subsection{Provenance Database}
Let $\mathcal{D} = \{\mathbf{r}_1, \mathbf{r}_2, \dots, \mathbf{r}_N\}$ represent the provenance database of size $N$, where each record $\mathbf{r}_i$ captures:
\begin{itemize}
    \item A \emph{process ID}, $p_i$.
    \item A set of \emph{events}, $\{e_{i1}, e_{i2}, \dots \}$.
 
\end{itemize}
These records collectively define a high-level, system-wide trace that can be converted into textual descriptions for embedding extraction.
\subsection{Textual Representation and LLM Embeddings}
To harness LLM-based embeddings, we convert each record $\mathbf{r}_i$ into a short sentence $s_i$. For instance, if $\mathbf{r}_i$ corresponds to a process with events $[ \text{open}, \text{write}, \text{send} ]$, we map it to:
$
\label{eq:text-rep}
    s_i = \text{``Process } p_i \text{ has event open and event write and event send.''}
$
\noindent \paragraph{Embedding Extraction.} We apply a pre-trained LLM, denoted generically by a function
$
    f_{\text{LM}}(\cdot): \text{Sentence} \;\mapsto\; \mathbb{R}^d,
$
which transforms the textual sentence $s_i$ into a $d$-dimensional dense \emph{embedding} $\mathbf{x}_i$. In our study, we experiment with several LLMs:
\begin{enumerate}
    \item BERT (Bidirectional Encoder Representations from Transformers) : Learns bidirectional context via masked language modeling. 
    \item DistilBERT : A distilled (compressed) variant of BERT with fewer parameters. It is produced via knowledge distillation, maintaining a good balance between accuracy and computational efficiency.
    \item {ALBERT} (A Lite BERT): Reduces model size through parameter sharing and factorized embeddings.
    \item {RoBERTa}  (Robustly Optimized BERT): An optimized variant of BERT trained with longer sequences and dynamic masking.
    \item {MiniLM} : Compresses self-attention to achieve a small yet effective model.
\end{enumerate}
Each model outputs a high-dimensional vector (e.g., 768 dimensions) encapsulating the process’s behavior in semantic form. By testing multiple LLMs, our framework evaluates how architectural differences (e.g., distillation, parameter sharing, encoder-decoder) impact the detection of stealthy APT.
Hence, for record $i$ we have the embedding: $\mathbf{x}_i = f_{\text{LM}}(s_i) \;\in\; \mathbb{R}^d.$ 
\subsection{Anomaly Detection with AutoEncoders}

Autoencoders (AEs) are unsupervised learning models that aim to encode input data into a compressed latent representation and reconstruct the original input from it. They are widely used for anomaly detection, where high reconstruction errors indicate data that deviates from the normal distribution. In our \emph{APT-LLM} framework, we extend a baseline AutoEncoder (AE) to its variants---Variational Autoencoders (VAEs) and Denoising Autoencoders (DAEs)---and introduce the use of attention mechanisms to enhance their performance. The motivation behind using a Variational Autoencoder (VAE) is to introduce a probabilistic latent space, enabling the model to better capture the underlying data distribution and generalize beyond reconstruction, while a Denoising Autoencoder (DAE) is designed to improve robustness by learning to reconstruct clean inputs from noisy data; both extend the baseline Autoencoder (AE), which focuses solely on deterministic compression and reconstruction without addressing probabilistic modeling (VAE) or noise resilience (DAE). 

\subsubsection{Baseline Autoencoder (AE)}

The baseline \textit{autoencoder} consists of two main components:
\begin{itemize}
    \item {Encoder} $E(x; \theta_E)$: Maps the input LLM embedding $x \in \mathbb{R}^d$ to a latent representation $z \in \mathbb{R}^k$, where $k \ll d$:
    $z = E(x; \theta_E) = f(W_E x + b_E)$ where $f(\cdot)$ is a nonlinear activation function, and $\theta_E = \{W_E, b_E\}$ are the encoder parameters.
    \item {Decoder} $D(z; \theta_D)$: Reconstructs the input $x$ from $z$: $\hat{x} = D(z; \theta_D) = g(W_D z + b_D)$ where $g(\cdot)$ is another nonlinear activation function, and $\theta_D = \{W_D, b_D\}$ are the decoder parameters.
\end{itemize}

The objective of AE is to minimize the \textit{reconstruction loss}: $\mathcal{L}_{\text{AE}} = \|x - \hat{x}\|^2$. This ensures that $z$ captures the most relevant features of $x$ while enabling accurate reconstruction.

\subsubsection{Variational Autoencoder (VAE)}

Unlike AE, \textit{VAE} introduces a probabilistic approach to the latent space by modeling $z$ as a random variable:
$z \sim \mathcal{N}(\mu, \sigma^2)$. The encoder outputs the mean $\mu$ and log-variance $\log \sigma^2$, which parameterize the latent Gaussian distribution:
$\mu, \log \sigma^2 = E(x; \theta_E)$. The decoder reconstructs $x$ by sampling from the latent distribution using the \textit{reparameterization trick}:$
z = \mu + \sigma \odot \epsilon, \quad \epsilon \sim \mathcal{N}(0, I)
$. The VAE objective combines the Reconstruction Loss: $\mathcal{L}_{\text{recon}} = \|x - \hat{x}\|^2$, and KL Divergence that regularizes the latent space to follow a standard Gaussian prior: $
    \mathcal{L}_{\text{KL}} = D_{\text{KL}}\big(\mathcal{N}(\mu, \sigma^2) \| \mathcal{N}(0, I)\big)
    $.
The total loss is:
$
\mathcal{L}_{\text{VAE}} = \mathcal{L}_{\text{recon}} + \beta \mathcal{L}_{\text{KL}}
$
where $\beta$ controls the weight of the KL divergence term.

\subsubsection{Denoising Autoencoder (DAE)}

\textit{DAE} extends AE by introducing \textit{noise} into the input data during training. The noisy input $\tilde{x}$ is generated by adding Gaussian noise or masking random features:
$
\tilde{x} = x + \eta, \quad \eta \sim \mathcal{N}(0, \sigma^2)
$. The encoder maps $\tilde{x}$ to a latent representation:
$
z = E(\tilde{x}; \theta_E)
$. The decoder reconstructs the original clean input $x$:
$
\hat{x} = D(z; \theta_D)
$. The loss function minimizes the reconstruction error between $x$ and $\hat{x}$:
$
\mathcal{L}_{\text{DAE}} = \|x - \hat{x}\|^2
$. This encourages the model to learn robust representations that ignore irrelevant noise.

\subsubsection{Attention Mechanisms}

To enhance the representational power of AE, VAE and DAE autoencoders, we incorporate \textit{self-attention layers} into the encoder. Self-attention enables the model to learn dependencies between different features, allowing it to focus on the most relevant aspects of the input.

Given an input $X \in \mathbb{R}^{n \times d}$ (the LLM embeddings), the attention mechanism computes:
$
\text{Attention}(Q, K, V) = \text{softmax}\left(\frac{Q K^T}{\sqrt{d_k}}\right)V
$
where $Q$, $K$, and $V$ are the query, key, and value matrices, respectively, derived from $X$. Multi-head attention extends this by using multiple attention mechanisms in parallel:
$
\text{MultiHead}(Q, K, V) = \text{Concat}(\text{head}_1, \dots, \text{head}_h)W^O
$. Integrating attention layers allows the encoder to capture complex relationships within LLM embeddings, improving anomaly detection performance.

\subsection{Experimental Plan}
During training, we use predominantly normal data, ensuring the autoencoders (AE, DAE, VAE) learn a compact representation of typical (non-APT) process embeddings.

To evaluate the different architectures in APT-LLM, we compare their effectiveness in anomaly detection using embeddings generated from LLMs (BERT, ALBERT, RoBERTa, DistilBert and MiniLM). Each model is trained on embeddings from normal process-action traces and tested on both normal and anomalous samples. Performance is measured using:
\begin{itemize}
    \item {Reconstruction Error}: Distribution of reconstruction errors for normal vs. anomalous data.
    \item {Anomaly Detection Metrics}: AUC-ROC.

\end{itemize}

The results will highlight the relative strengths of each architecture and the impact of attention layers on anomaly detection.

\subsubsection{Anomaly Scoring}
After training on normal embeddings, the reconstruction error serves as an \emph{anomaly score}.
$
\label{eq:recon-error}
    r_i = \left\lVert \mathbf{x}_i - \hat{\mathbf{x}}_i \right\rVert_2^2
$
If $r_i$ exceeds a threshold $\tau$, the record is flagged as a potential APT:
$
\label{eq:threshold}
    \text{flag}(\mathbf{x}_i) = \begin{cases}
    1, & \text{if } r_i > \tau, \\
    0, & \text{otherwise}.
    \end{cases}
$. \\
Choosing $\tau$ can be based on a percentile of $r_i$ among known normal samples or tuned via validation metrics.

\subsection{Workflow Summary}
\begin{enumerate}
\item {Data Collection}: Extract a provenance database $\{\mathbf{r}_1,\dots,\mathbf{r}_N\}$ containing process actions, system calls, and other metadata.
\item {Sentence Generation}: Map each record $\mathbf{r}_i$ to a sentence $s_i$.
\item {LLM Embedding}: Compute $\mathbf{x}_i = f_{\text{LM}}(s_i)$ for each record, where $f_{\text{LM}}$ is one of our chosen LLMs (BERT, DistilBERT, ALBERT, RoBERTa, or MiniLM).
\item {Noise Injection \& Encoding}: Obtain $\tilde{\mathbf{x}}_i$ by adding noise to $\mathbf{x}_i$, then encode $\tilde{\mathbf{x}}_i$ to latent vector $\mathbf{z}_i$.
\item {Training}: Use normal data and train AE, VAE and DAE.
\item {Decoding \& Reconstruction}: Decode $\mathbf{z}_i$ back to $\hat{\mathbf{x}}_i$, training on normal embeddings to minimize $\left\lVert \mathbf{x}_i - \hat{\mathbf{x}}_i \right\rVert_2^2$.
\item {Anomaly Detection}: For new (test) samples, compute reconstruction errors $r_i$. If $r_i > \tau$, label the sample as anomalous.
\end{enumerate}

\section{Experimental settings, results and analysis}
\subsection{Provenance Datasets}
The datasets used in this study are derived from DARPA's \verb|Transparent Computing TC| program\footnote{https://www.darpa.mil/program/transparent-computing}, processed through the ADAPT\footnote{https://gitlab.com/adaptdata} (Automatic Detection of Advanced Persistent Threats) project's ingester. These datasets span four OS —Android, Linux, BSD, and Windows— and include two attack scenarios: Pandex and Bovia~\cite{BerradaCBMMTW20},\cite{Benabderrahmane21},\cite{DBLP:journals/fgcs/BenabderrahmaneHVCR24}. Provenance graph data is processed into graph databases, followed by data integration and deduplication, resulting in Boolean-valued datasets that represent various aspects of system process behaviors. Each dataset row corresponds to a process and is encoded as a Boolean vector, where a value of 1 indicates the presence of a specific attribute or event. Five data aspects are constructed for each OS and attack scenario: ProcessEvent (PE), which captures event types performed by processes; ProcessExec (PX), representing executable names starting the processes; ProcessParent (PP), denoting executables of parent processes; ProcessNetflow (PN), detailing IP addresses and port names accessed by processes; and ProcessAll (PA), a union of all attribute sets. With two scenarios, four OS, and five data aspect, a total of 40 datasets are created. Table \ref{datatable} summarizes them, where the last column illustrates their highly imbalanced nature, with APT attacks constituting as little as 0.004\% of the data in some cases.
\begin{table*}
%\vspace{-6 em}
 \captionsetup{singlelinecheck=false}
\centering
\small
\resizebox{0.99\textwidth}{!}{
\begin{tabular}{|l|l||l|l|l|l|l|l|l|l|}
\hline & Scenario & Size& $PE$   & $PX$  & $PP$  & $PN$     & $PA$  & $nb\_attacks$    & $\%\frac{nb\_attacks}{nb\_processes}$     \\ \hline \hline
BSD    & 1 &288 MB &76903 / 29  & 76698 / 107  & 76455 / 24  & 31 / 136  & 76903 / 296 & 13&0.02\\  
    & 2 &1.27 GB &224624 / 31  &224246 / 135  & 223780 / 37  & 42888 / 62 &  224624 / 265      & 11&0.004\\ \hline
Windows & 1 &743 MB & 17569 / 22    &  17552 / 215  &   14007 / 77        &   92 / 13963      & 17569 / 14431& 8&0.04\\  
   & 2 &9.53 GB& 11151 / 30    &  11077 / 388  & 10922 / 84  & 329 / 125      &  11151 / 606    &8&0.07\\ \hline
Linux  & 1 &2858 MB &247160 / 24 & 186726 / 154 & 173211 / 40 & 3125 / 81 & 247160 / 299  &25&0.01\\
    & 2 &25.9 GB &282087 / 25 & 271088 / 140 & 263730 / 45 &6589 / 6225 &  282104 / 6435      &46&0.01\\ \hline
Android& 1 &2688 MB&102 / 21     &102 / 42&0 / 0&8 / 17& 102 / 80&9&8.8\\
&2 &10.9 GB&12106 / 27     &12106 / 44&0 / 0&4550 / 213&12106 / 295 &13&0.10\\ \hline
\end{tabular}
}

\caption{Experimental datasets of DARPA's TC program used in our study. A dataset entry (columns 4 to 8) is described by a number of rows (processes) / number of columns (attributes). For instance, with ProcessAll (PA) obtained from the second scenario using Linux, the dataset has 282104 rows and 6435 attributes with 46 APT attacks (0.01\%). }
   % \vspace{-5mm} % Adjust value as needed
 \label{datatable}
\end{table*}

\subsection{Illustrative Examples of LLM Embeddings}

To explain the embedding generation process, we convert a sample process trace into textual representations and pass it through the LLMs. Each model uses its unique architecture and training strategy but outputs dense embeddings that capture the semantic meaning of the input.

%\subsection{Example Process Trace}
Suppose a process (id:123) performs the following events:
\begin{itemize}
    \item Open a file 
    \item Write to the file
    \item Send network data
\end{itemize}
This trace is represented as a sentence: 
\begin{quote}
    \texttt{``Process 123 has event open a file and event write to the file and event send network data.''}
\end{quote}
We pass this sentence into the different LLMs, which tokenize it, process it through their architectures, and produce dense embeddings.\\
Initially, BERT runs a tokenization process by splitting the input sentence into tokens, e.g., \texttt{[CLS, Process, 123, has, event, open, .., event, write, .., event, send, .., [SEP]]}. Each token is converted into a numeric ID. Then BERT uses bidirectional attention to learn the context of each token based on both preceding and following tokens. For example, the embedding for \texttt{event open a file} incorporates its relationships to \texttt{event write to the file} and \texttt{event send network data}. The final output is a 768-dimensional vector for each token. To get a single vector for the sentence, we use mean pooling over all token embeddings or select the embedding of the \texttt{[CLS]} token. ALBERT on the other hand-side reduces model size by factorizing embedding parameters, and cross-layer parameter sharing. Similar to BERT, ALBERT tokenizes the input and generates embeddings using bidirectional attention. Despite its smaller size, ALBERT captures semantic relationships effectively. RoBERTa improves BERT by removing the next-sentence prediction task. It processes the input sentence similarly to BERT but achieves richer embeddings due to better training optimizations. Each token is represented by a 1024-dimensional vector (for the large model), and sentence embeddings are derived through pooling. DistilBERT is a lighter, faster version of BERT, trained using knowledge distillation to retain 97\% of BERT’s performance with ~40\% fewer parameters. The input is tokenized and passed through a smaller Transformer network, reducing computation time while still generating high-quality embeddings. Sentence embeddings are derived as 768-dimensional vectors, like BERT, but computed more efficiently. MiniLM compresses the attention mechanism in Transformers, significantly reducing the model size while maintaining strong performance on semantic tasks. MiniLM tokenizes the input and uses a highly compressed Transformer network to generate embeddings. Its efficiency makes it ideal for real-time applications. Embeddings are 384-dimensional vectors, smaller than BERT, making it computationally lightweight.

\subsection{Comparison of LLMs}
Using these multiple LLMs  allows us to compare:
\begin{itemize}
    \item {Embedding Quality:} Larger models like RoBERTa might capture richer semantic nuances, while smaller models like DistilBERT and MiniLM provide efficiency.
    \item {Detection Performance:} We evaluate how different embeddings impact anomaly detection performance (e.g., reconstruction error, AUC).
\end{itemize}
\subsection{Embeddings visualization}
\begin{figure}[ht]
 \captionsetup{singlelinecheck=false}
    \centering
    % First Row
    \begin{subfigure}[t]{0.35\linewidth}
        \centering
        \includegraphics[width=\linewidth]{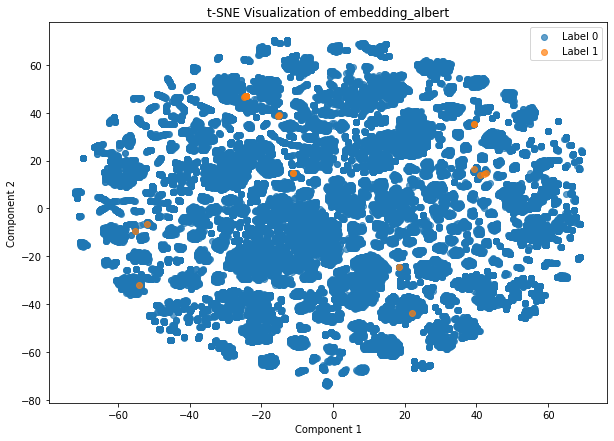}
        \caption{ALBERT}
    \end{subfigure}
    \begin{subfigure}[t]{0.35\linewidth}
        \centering
        \includegraphics[width=\linewidth]{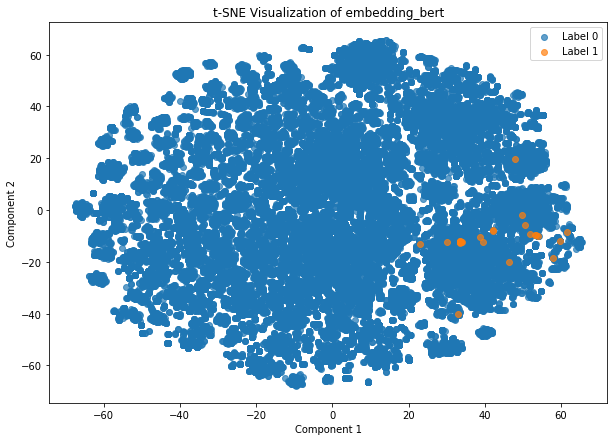}
        \caption{BERT}
    \end{subfigure}
    \begin{subfigure}[t]{0.35\linewidth}
        \centering
        \includegraphics[width=\linewidth]{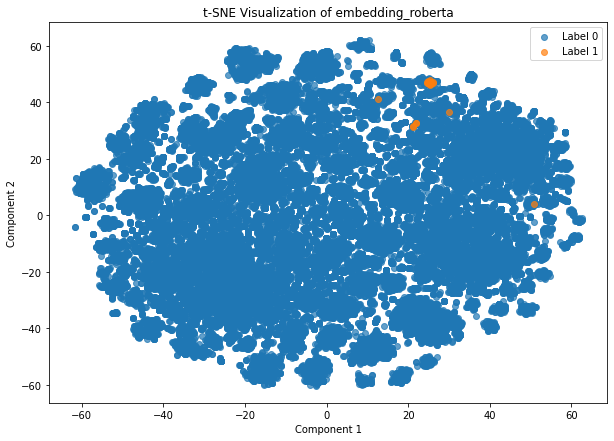}
        \caption{RoBERTa}
    \end{subfigure}
    
    % Second Row
   % \vspace{0.5cm} % Add spacing between rows
    \begin{subfigure}[t]{0.35\linewidth}
        \centering
        \includegraphics[width=\linewidth]{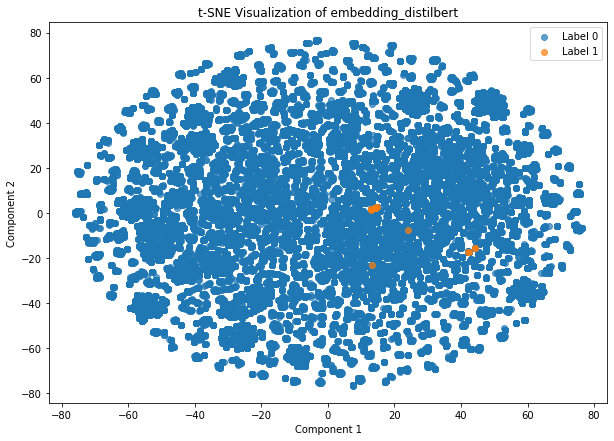}
        \caption{DistilBERT}
    \end{subfigure}
    \begin{subfigure}[t]{0.35\linewidth}
        \centering
        \includegraphics[width=\linewidth]{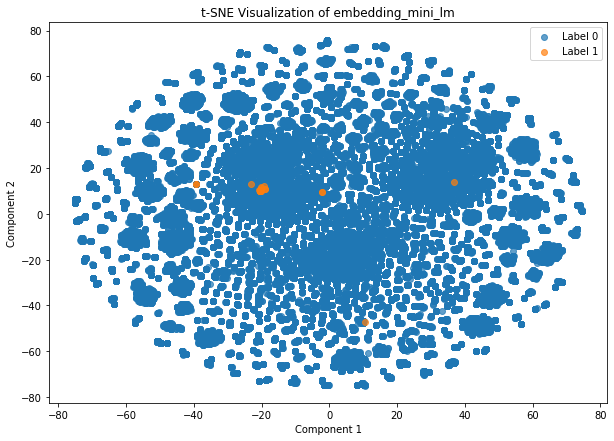}
        \caption{MiniLM}
    \end{subfigure}

    \caption{T-SNE Visualizations of Embeddings Using Different LLMs. Blue points (label 0) represent normal data, whereas orange points (label 1) represent anomalies. In this example, data belongs to PE dataset of Linux OS and Bovia scenario.}
    \label{fig:tsne-llms}
\end{figure}

Figure \ref{fig:tsne-llms} provides T-SNE visualizations of the embeddings generated using five large language models (LLMs), showcasing the projection of normal data (blue points, label 0) and anomalies (orange points, label 1). In this example, the data points correspond to the PE data set of the Linux OS under the Bovia attack scenario. The embeddings highlight how LLM-based representations capture nuanced differences between normal and anomalous patterns. We generated these T-SNE visualizations for the entire collection of datasets for providing a clear clustering of normal points while isolating anomalies. This visualization demonstrates the potential of LLMs in embedding provenance trace data for effective anomaly detection.
%%%%%%%%%%%%%%%%%%%%%%%
\subsection{AutoEncoders training and reconstruction errors}
\begin{figure}
 \captionsetup{singlelinecheck=false}
    \centering
    \includegraphics[width=0.65\linewidth]{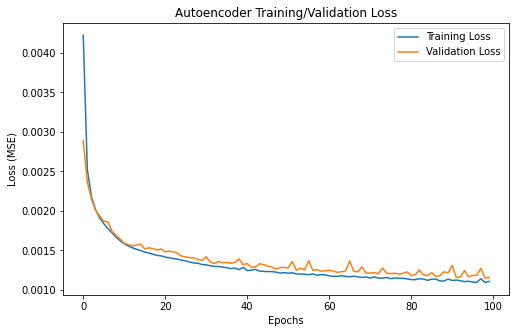}
    \caption{Training and Validation Loss Curve for the Autoencoder: The figure shows the decrease in training and validation loss (MSE) over 100 epochs, demonstrating the model's convergence and generalization during training.}
     %   \vspace{-4mm} % Adjust value as needed
    \label{fig:ErrorAE}
\end{figure}

The embedding vectors generated in the previous step with the five LLMs are used to train AE, VAE and DAE.
Figure \ref{fig:ErrorAE} shows the training and validation loss curves for the baseline autoencoder over 100 epochs, using the previous dataset (PE, Linux, Bovia). Both curves exhibit a steep decline during the initial epochs, indicating rapid learning and optimization. After approximately 20 epochs, the losses stabilize and decrease gradually, suggesting convergence of the model. The training loss remains slightly lower than the validation loss, highlighting a well-generalized model with minimal overfitting. The consistent alignment between the two curves over time reflects the autoencoder's ability to learn meaningful representations of the input data while maintaining robustness on unseen validation samples.
%%%%%%
\begin{figure}
 \captionsetup{singlelinecheck=false}
    \centering
    \includegraphics[width=0.65\linewidth]{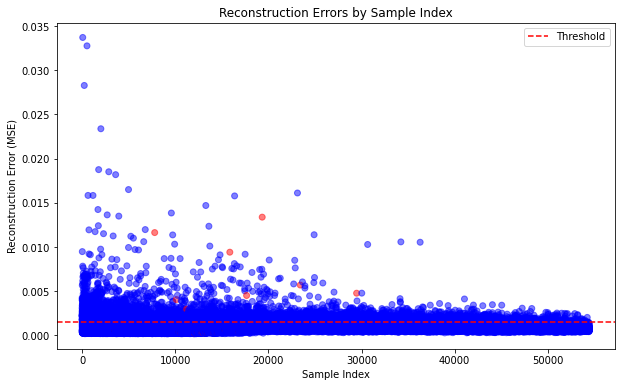}
    %}
    \caption{Scatter Plot of the AutoEncoder Reconstruction Errors by Sample Index: The figure illustrates the reconstruction errors (MSE) for each sample, with the red dashed line indicating the anomaly detection threshold. Points above the threshold represent detected anomalies.}
    \label{fig:scatterplot}
\end{figure}
Figure \ref{fig:scatterplot} presents a scatter plot of reconstruction errors (MSE) from the baseline autoencoder. Each point represents a sample, with anomalies (red) identified when errors exceed the red dashed threshold. Most data points fall below this threshold, indicating normal samples, while a few outliers exceed it, demonstrating the effectiveness of reconstruction error for anomaly detection.\\
\begin{figure}
 \captionsetup{singlelinecheck=false}
\centering
\includegraphics[width=0.5\linewidth]{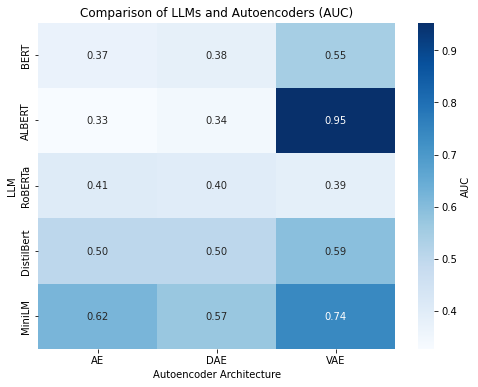}
    \caption{Heatmap of AUC Scores for LLM and Autoencoder Combinations: The figure shows the performance of five language models (LLMs) paired with three autoencoder architectures (AE, DAE, VAE) on anomaly detection tasks, with darker shades indicating higher AUC scores. Data belongs to PE dataset of Linux OS and Bovia scenario.}
       % \vspace{-6mm} % Adjust value as needed
    \label{fig:Heatmap}
\end{figure}

%%%%%%%%%%%%
Given the combination of 5 LLMs and 3 autoencoder architectures (AE, VAE, and DAE), we conducted extensive experiments by systematically testing all configurations (5$\times$3) to identify the best-performing model for anomaly detection. For each combination, we generated ROC curves and calculated the corresponding AUC scores to comprehensively evaluate their performance.
The heatmap in Figure \ref{fig:Heatmap} provides an overview of the AUC scores achieved by combinations of the five language models, and the autoencoder architectures (AE, DAE, and VAE) for the PE Linux Bovia dataset. ALBERT paired with VAE significantly outperforms other combinations, achieving the highest AUC score of 0.95, highlighting its ability to capture and reconstruct patterns effectively in this architecture. MiniLM also exhibits consistent performance across all autoencoder types, with its best result (AUC = 0.74) observed when combined with the VAE, demonstrating its lightweight yet robust capabilities. While DistilBERT shows moderate performance, peaking with VAE (AUC = 0.59), BERT and RoBERTa generally underperform except for slight improvements in certain configurations. The analysis underscores the importance of selecting the right pairing of LLMs and autoencoders to optimize anomaly detection performance, with ALBERT-VAE emerging as the clear winner for the considered dataset.

\begin{figure}[h!]
 \captionsetup{singlelinecheck=false}
    \centering
    % First figure
  %  \begin{subfigure}[b]{0.45\textwidth}
   %     \includegraphics[width=0.85\textwidth]{ROC_Linux_Bovia_PE_MuliLLM_v4.png}
   %     \caption{Combining five LLMs and 3 AutoEncoder architectures.}
    %    \label{fig:figure1}
    %\end{subfigure}
    %\hfill
    % Second figure
  %  \begin{subfigure}[b]{0.45\textwidth}
        \includegraphics[width=0.55\linewidth]{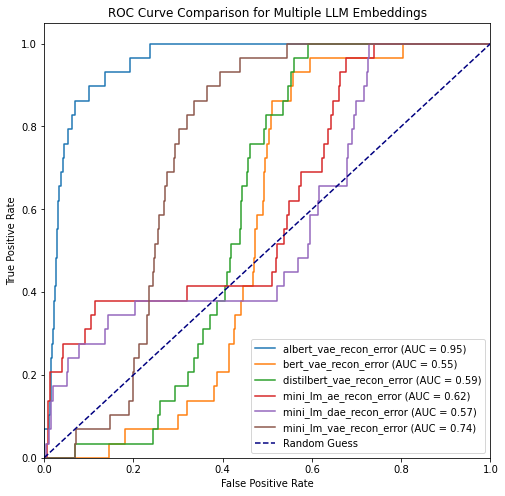}
   %     \caption{Best performing models.}
       % \label{fig:figure2}
   % \end{subfigure}
    \caption{ROC Curve Comparison for the Best performing models (PE dataset of Linux OS and Bovia scenario).}
    \label{fig:ROCcombined}
\end{figure}

\begin{figure}[h!]
 \captionsetup{singlelinecheck=false}
    \centering
    \includegraphics[width=1\linewidth]{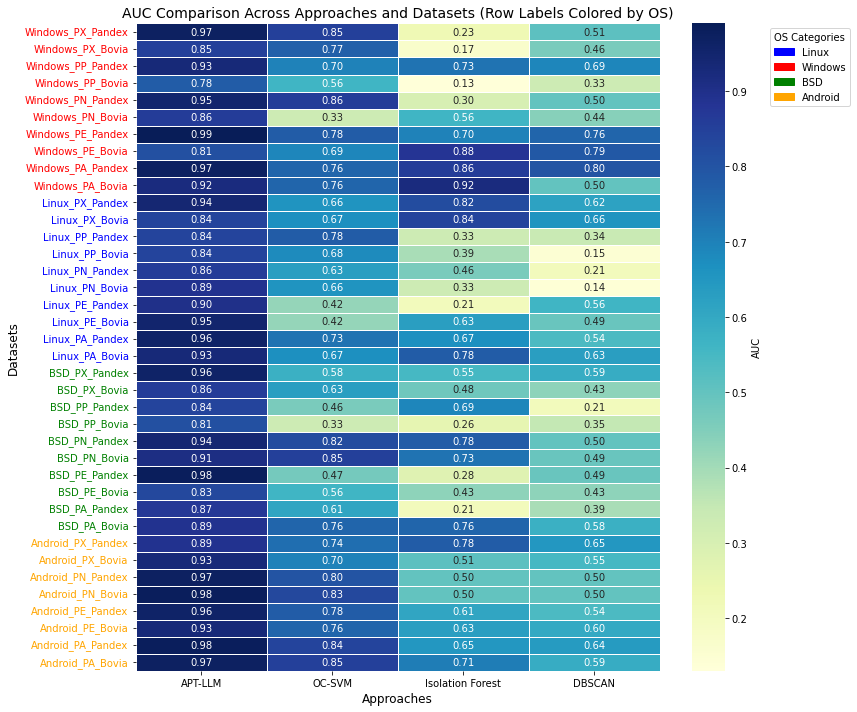}
    \caption{AUC heatmap comparing the performance of APT-LLM, OC-SVM, Isolation Forest, and DBSCAN across datasets from multiple OS (Windows, Linux, BSD, and Android) and provenance trace aspects. APT-LLM consistently achieves higher AUC scores, demonstrating its superior anomaly detection capabilities}
       % \vspace{-5mm} % Adjust value as needed
    \label{fig:auc-heatmap}
\end{figure}

Figure \ref{fig:ROCcombined} illustrates the ROC curves of the top-6 best-performing combinations identified from the heatmap, highlighting their superior anomaly detection capabilities.
The ROC curves provide a comparative analysis of the performance of the different combinations of LLMs paired with three autoencoder architectures. Each curve represents the true positive rate (TPR) against the false positive rate (FPR) for a specific LLM-autoencoder pairing, with the area under the curve (AUC) indicating the overall performance. \\
We generated heatmaps of the AUC scores for all combinations of LLMs and autoencoder architectures across the remaining datasets (5$\times$3$\times$40), and for each dataset, we identified the combination yielding the maximum AUC to serve as the baseline \emph{APT-LLM} for comparison with existing anomaly detection methods. Indeed, for each dataset, the best-performing combination (LLM × AE) was evaluated against three classical anomaly detection approaches—OC-SVM, DBSCAN, and Isolation Forest. The resulting AUC scores are presented in Heatmap \ref{fig:auc-heatmap}.

%%%%%%%%
The results reveal significant differences in the AUC performance of various anomaly detection methods—APT-LLM, OC-SVM, Isolation Forest, and DBSCAN—across datasets spanning multiple OS (Windows, Linux, BSD, and Android) and provenance trace aspects. APT-LLM consistently outperforms the baseline methods in most datasets, achieving the highest AUC scores. For example, in PX Windows  Pandex, APT-LLM scores 0.97, compared to OC-SVM (0.85), Isolation Forest (0.23), and DBSCAN (0.51). Baseline methods such as OC-SVM show competitive performance in a few cases, particularly in BSD datasets, while Isolation Forest and DBSCAN often show weaker results. These findings highlight the robustness and superior detection capabilities of the APT-LLM framework across diverse datasets and scenarios.

\section{Conclusion}
In this paper, we presented \emph{APT-LLM}, an embedding-based anomaly detection framework that leverages large language models and autoencoders to identify advanced APT cyber threats. By transforming process behaviors into textual representations, our approach introduces a novel perspective for anomaly detection tasks. Experimental results highlight significant improvements in APT detection performance, particularly in highly imbalanced scenarios, demonstrating the effectiveness and potential of LLM-based embeddings in advancing cybersecurity practices.

\bibliographystyle{IEEEtran}
\bibliography{references}

\end{document}